\documentclass[twocolumn,prd,
 reprint,
nofootinbib,
amsmath,amssymb,
aps
]{revtex4}

\usepackage{comment}
\usepackage{feynmp}
\usepackage[pdftex]{graphicx}
\usepackage{dcolumn}

\usepackage{dsfont}
\usepackage{graphicx}
\usepackage{dcolumn}
\usepackage{bm}
\usepackage{hyperref}

\begin{document}

\title{Stochastic determination of matrix determinants}

\author{Sebastian Dorn\footnote{sdorn@mpa-garching.mpg.de} and Torsten A. En{\ss}lin}
\affiliation{Max-Planck-Institut f\"ur Astrophysik, Karl-Schwarzschild-Str.~1, D-85748 Garching, Germany\\
Ludwigs-Maximilians-Universit\"at M\"unchen, Geschwister-Scholl-Platz 1, D-80539 M\"unchen, Germany}
\date{\today}

\begin{abstract}
Matrix determinants play an important role in data analysis, in particular when Gaussian processes are involved. Due to currently exploding data volumes,
linear operations -- matrices -- acting on the data are often not accessible directly but are only represented indirectly in form of a
computer routine. Such a routine implements the transformation a data vector undergoes under matrix multiplication. 
While efficient probing routines to estimate a matrix's diagonal or trace, based solely on such computationally affordable matrix-vector multiplications,
are well known and frequently used in signal inference, there is no stochastic estimate for its determinant. 
We introduce a probing method for the logarithm of a determinant of a linear operator. 
Our method rests upon a reformulation of the log-determinant by an integral representation and the transformation of the involved terms into stochastic
expressions. This stochastic determinant determination enables large-size applications in Bayesian inference, in particular evidence calculations,
model comparison, and posterior determination.

\end{abstract}

\maketitle
\section{Motivation}\label{sec:motivation}
Current and future physical observations generate huge data streams to be analyzed. Particle physics, biophysics, astronomy, and cosmology are representatives of 
current scientific fields of interest that are undergoing a revolution driven by increasing data volume. Typical large data sets in cosmology are, for instance,
observations of the cosmic microwave background \cite{2013ApJS..208...20B,2014A&A...571A..16P}
as well as of the large-scale structure \cite{2000AJ....120.1579Y,2007ApJS..170..377S,2008ASPC..399..115H} 
as they are often wide- or all-sky observations carried out by telescopes with remarkable resolution. 
In order to extract information about the universe or physics in general, Bayesian inference methods becomes more and more frequently used as their large computational
demands become more feasible thanks to technology developments. The signal of interest to be extracted from data could be 
almost everything, ranging from just a single parameter (e.g.,~the level of local non-Gaussianity of the cosmic microwave 
background \cite{2005ApJ...634...14K,paper1})
to a full four-dimensional reconstruction of the structure growth in the universe \cite{2008MNRAS.389..497K,2013MNRAS.432..894J}.
Such ambitious Bayesian analyses often invoke linear transformations of the data or of estimated signal vectors.

The size of the involved data and signal spaces often bans the explicit representation of matrices acting on these spaces by their individual matrix elements.
A prominent example appearing in many analyses is, for instance,
the covariance matrix of a multivariate Gaussian distribution of a vector valued quantity, which describes the two-point correlation structure of the said quantity.
Due to their large dimensions such matrices are often only representable by a computer routine, which implements the application of the matrix to a vector
without storing or even calculating the individual matrix elements. Such routines often invoke fast Fourier transformations and other efficient operations,
which in combination render nonsparse matrices into easily computable basis systems. We refer to such a matrix as an \textit{implicit matrix}.
For instance, calculating the model evidence often requires calculating determinants of such matrices. 
This work provides an efficient way to numerically calculate determinants given only by an implicit matrix representation.

The remainder of this work is organized as follows: In Sec.~\ref{sec:main} we introduce the formalism of the stochastic estimation of an implicit matrix and present
two numerical examples. Section \ref{sec:applications} provides a perspective of possible applications in science. Results are summarized in Sec.~\ref{sec:summary}.

\section{Probing the log-determinant of an implicit matrix}\label{sec:main}
\subsection{Formalism}
Let $A=(a_{ij})\in \mathds{C}^{n\times n}$ be an implicitly defined, complex-valued, square matrix of order $n$. 
Implicitly means that the particular entries of the matrix are not accessible, for instance,
if dealing with large data sets, where an explicit storage of $A$ might exceed the memory of the computer. However, the action of the matrix as a linear operator is 
assumed to be known and given by a computer routine implementing the mapping $x\mapsto Ax$.

Motivated by applications in science and statistics (Secs.~\ref{sec:motivation} and \ref{sec:applications}), in particular by signal reconstruction techniques
and model comparison in astronomy and cosmology, where the determinant of a covariance matrix is required (Sec.~\ref{sec:applications}), we constrain
the variety of different types of matrices by requesting that the matrix $A$ of interest is either
\textit{weak diagonal dominant} or \textit{Hermitian positive definite}. The term \textit{weak diagonal dominant} is defined by 
\begin{equation}
 |a_{ii}|\geq \sum_{i\neq j}|a_{ij}|~~~\forall i, 
\end{equation}
while \textit{Hermitian positive definite} means
\begin{equation}
 A^\dag = A ~~~\mathrm{and}~~~x^\dag A x >0 ~~~\forall x\in\mathds{C}^n{\backslash}\{0\}
\end{equation}
with $\dag$ denoting the adjoint. 

The diagonal and the trace of an implicit matrix can be obtained by exploiting common probing routines 
\cite{2012PhRvE..85b1134S,2011arXiv1105.5256A,doi:10.1080/03610918908812806,Bekas20071214}. A stochastic estimate of the diagonal of the linear
operator $A$ is given by
\begin{equation}\label{eq:diag}
 \mathrm{diag}(A) = \left\langle \xi \star A\xi \right\rangle_{\{\xi\}} \approx \frac{1}{M}\sum_{i=1}^M \xi_i \star A\xi_i,
\end{equation}
where $\star$ denotes a componentwise product, $M=|\{\xi\}|$ the sample size, and $\left\langle \cdot \right\rangle_{\{\xi\}}$ the arithmetic mean over 
$\xi$ with $M\rightarrow \infty$. The probing vectors $\xi \in \mathds{C}^{n}$ are random variables, whose components $x~(x')$ fulfill the condition 
$\left\langle \xi_x\xi_{x'} \right\rangle_{\{\xi\}} = \delta_{xx'}$.
Analogously to the diagonal of an operator its trace can be probed by, e.g.,
\begin{equation}\label{eq:trace}
 \mathrm{tr}(A) = \left\langle \xi^\dag  A\xi \right\rangle_{\{\xi\}}.
\end{equation}

Recently, there have been investigations to improve these straightforward probing methods by exploiting Bayesian inference \cite{2012PhRvE..85b1134S}. This has 
been achieved by reformulating the process of stochastic probing of an operator's diagonal (trace) as a signal inference problem.
As a result, it requires 
fewer probes than the purely stochastic methods and thus can decrease the computational costs. With the phrase \textit{operator probing},
be it trace or diagonal probing, we subsequently refer to the entirety of probing methods in general.

The linear operator $A$ can be split into a diagonal matrix
$D\in \mathds{C}^{n\times n}$ and a matrix $N\in \mathds{C}^{n\times n}$, which contains the off-diagonal part of $A$, i.e.,
\begin{equation}\label{eq:split}
 A = D + N.
\end{equation}
We are now interested in the value of its determinant or of its log-determinant, $\Delta \equiv \ln[\mathrm{det}(A)]$. In case $A$
is mainly dominated by its diagonal (i.e. $ND^{-1} \ll \mathds{1}$ spectrally), a Taylor expansion of the log-determinant might be a reasonable approximation,
\begin{equation}\label{eq:approx}
\begin{split}
 \Delta =&~ \ln[\mathrm{det}(D+N)] \\
 =&~ \ln[\mathrm{det}(D)] + \mathrm{tr}\left[ N D^{-1}\right] + \mathcal{O}\left(\mathrm{tr}\left[ \left(ND^{-1}\right)^2\right]\right),
\end{split}
\end{equation}
which is sometimes feasible dealing with implicit operators, e.g.,~see Refs.~\cite{paper1,2014JCAP...06..048D} for recent applications in cosmic microwave background
physics. This approximation, however, breaks down when the relation $ND^{-1} \ll \mathds{1}$ (spectrally) is violated. 
In order to circumvent this problem we introduce 
the quantity
\begin{equation}
 \Delta(t) \equiv \ln[\mathrm{det}(D+t N)] 
\end{equation}
with the pseudotime parameter $t\in [0,1]$. For a sufficiently small $t$ the approximation of Eq.~(\ref{eq:approx}) becomes valid. This property can be used together
with a few mathematical manipulations (for details see Appendix~\ref{app:integral}) to obtain the formula 
\begin{equation}\label{eq:main_res}
\begin{split}
\Delta =& \int_0^1 dt~ \mathrm{tr}\left[ N\left( D + tN \right)^{-1}\right] + \Delta(0)\\
	=&~ \int_0^1 dt~\left\langle \xi^\dag  N\left( D + tN \right)^{-1} \xi \right\rangle_{\{\xi\}} + \Delta(0)
\end{split}
\end{equation}
that represents a stochastic estimate of the log-determinant of $A$ using operator probing. In particular, the following steps are required to 
evaluate Eq.~(\ref{eq:main_res}):
\begin{enumerate}
 \item Diagonal (operator-) probing to split $A$ into $$A = \underbrace{\mathrm{diag}(A)}_{\equiv D} + \underbrace{A - \mathrm{diag}(A)}_{\equiv N},$$
 \item an approach to invert $ D + tN$ in Eq.~(\ref{eq:main_res}), e.g.,~the \texttt{conjugate gradient} method \cite{CG},
 \item trace (operator-) probing to evaluate the 
 integrand
,
 \item a numerical integration method, e.g.,~applying \texttt{Simpson's} rule.
\end{enumerate}

It might immediately strike the eye of the reader that one recaptures the simple first-order Taylor-expanded version of the log-determinant, Eq.~(\ref{eq:approx}),
when dropping the pseudotime dependency of the integrand in Eq.~(\ref{eq:main_res}) by requesting $t=0$. This means that in case of dealing with diagonal 
dominant operators the value of the correct log-determinant might be received by a coarse numerical integration since the integrand close to $t=0$
already yields the main correction, which might decrease the computational costs, see Sec.~\ref{sec:num}. 

Equation (\ref{eq:main_res}) further represents the main result of this paper and can be regarded as a special case of calculating partition functions (see
Sec.~\ref{sec:applications} and Refs.~\cite{partition,Dickson2004}). Although the first line of it, the integral representation of the log-determinant,
was also, independently of our work, found by mathematicians 10 years ago \cite{du2005integral}, it is (to our knowledge) not known in the community of physics
or signal inference. The connection to stochastic estimators, however, is a novel way to evaluate the log-determinant of implicitly defined matrices
that enables previously impossible calculations, see Sec.~\ref{sec:applications}. 

\subsection{Numerical example}\label{sec:num}
\begin{figure}[ht]
\includegraphics[width=.8\columnwidth]{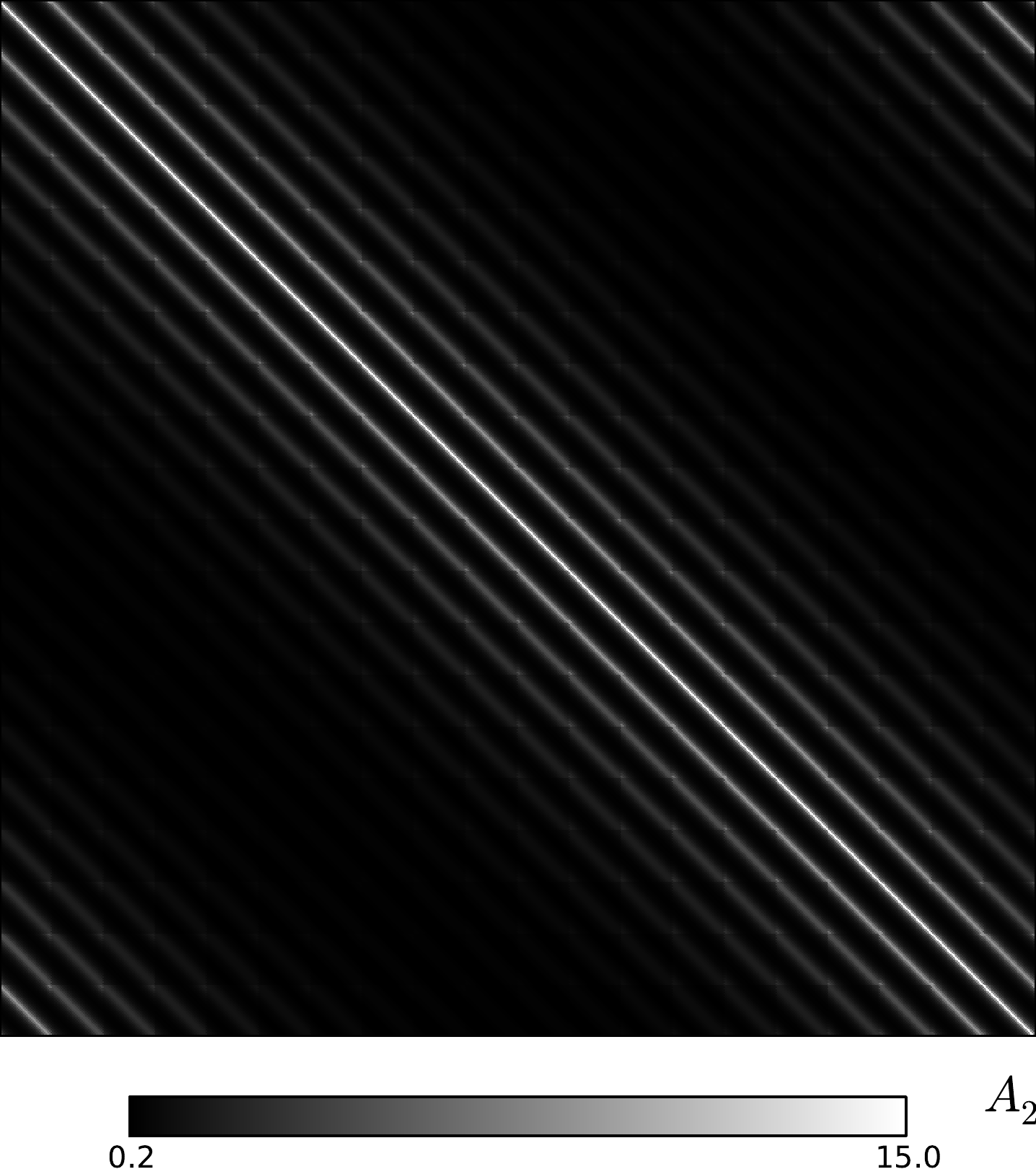}

\vspace{.1cm}
\includegraphics[width=.8\columnwidth]{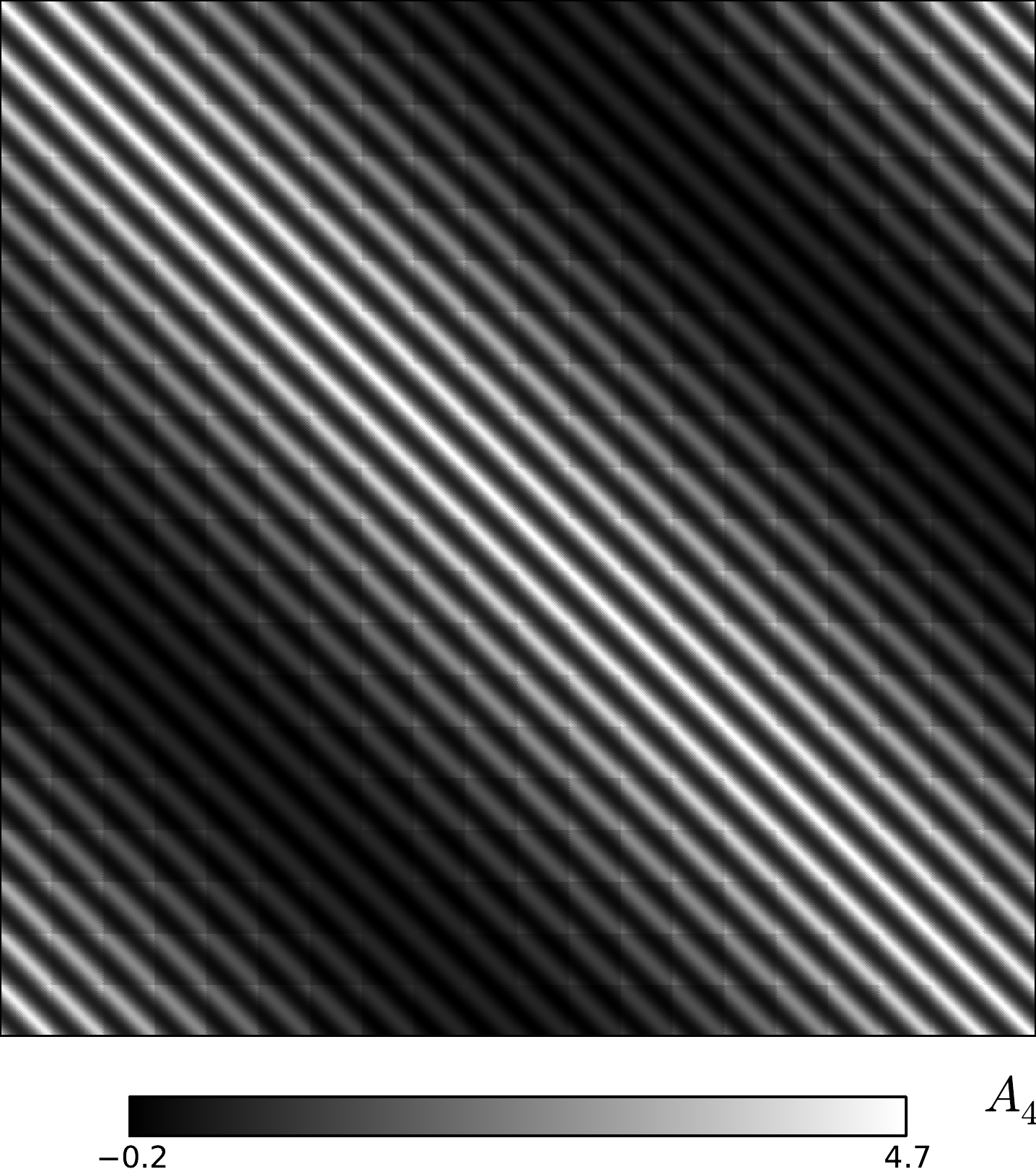}
\caption{Illustration of the matrices $A_2$ (top) and $A_4$ (bottom) in position space with linear color bars.}%
\label{fig:matricies}
\end{figure}

We address here a simple and also exactly solvable numerical example referring to (Bayesian) signal inference problems or, in general, statistical problems 
in physics (see Secs.~\ref{sec:evidence} and \ref{sec:post}), where the log-determinant of a covariance matrix $A$ is of interest. If we assume statistical
isotropy and homogeneity of a physical field, its covariance matrix can be parametrized by a so-called power spectrum. This is often a reasonable 
assumption\footnote{Referring to Bayesian evidence calculations such a matrix might be the prior or posterior covariance, see Sec.~\ref{sec:applications}
for details.}, e.g.,~in
astronomy and physical cosmology, when applying the cosmological principle.  In this case the covariance matrix becomes diagonal in Fourier space,
\begin{equation}
A_{kk'} = c_k \delta_{kk'},
\end{equation}
with respective Fourier modes $k,~k'$ and power spectrum $c_k$. It is straightforward to show that the position space representation of $A_{kk'}$, given by
$A_{xx'}=\mathcal{F}^{\dag}_{xk}A_{kk'}\mathcal{F}_{k'x'}$ with Fourier transformation $\mathcal{F}$, is nondiagonal
if and only if $c_k \neq \mathrm{const}~ \forall k$. In order to apply the stochastic estimator of the log-determinant we use two special forms
of the power spectrum, given by
\begin{equation}
c_k = \frac{1}{\left(1 + k\right)^\alpha}
\end{equation}   
with $\alpha$ set to 2 or 4. A value of $\alpha=2$ describes a mostly diagonal dominant matrix, whereas $\alpha=4$ exhibits a significant nondiagonal
structure in position space. 
To be precise, in the following we use a regular, two-dimensional, real-valued grid (over $\mathcal{T}^2$) of $n=20\times20$ pixels to represent our position space, resulting in a matrix $A$
consisting of $n\times n = 1.6\times10^5$ real numbers. See Fig.~\ref{fig:matricies} for an illustration thereof. 

For both matrices, which we refer to as $A_2$ and $A_4$, we apply Eq.~(\ref{eq:main_res}) given an explicit and implicit numerical implementation. 
For the explicit variant there also exist well-understood, precise numerical 
methods\footnote{See, for instance, the method described at \url{http://docs.scipy.org/doc/numpy/reference/generated/numpy.linalg.slogdet.html}, 
which is based on LU-factorization.} to calculate the determinant. Therefore, the numerical results of such a method can be regarded as our gold standard
and hence serve as a reference for the probing results. Henceforth we will refer to it using the subscript ``correct''. 
Both variants, the explicit and implicit implementation,
are realized using the tools of \textsc{NIFTy} \cite{2013AA...554A..26S}. 

\begin{figure}[ht]
\includegraphics[width=\columnwidth]{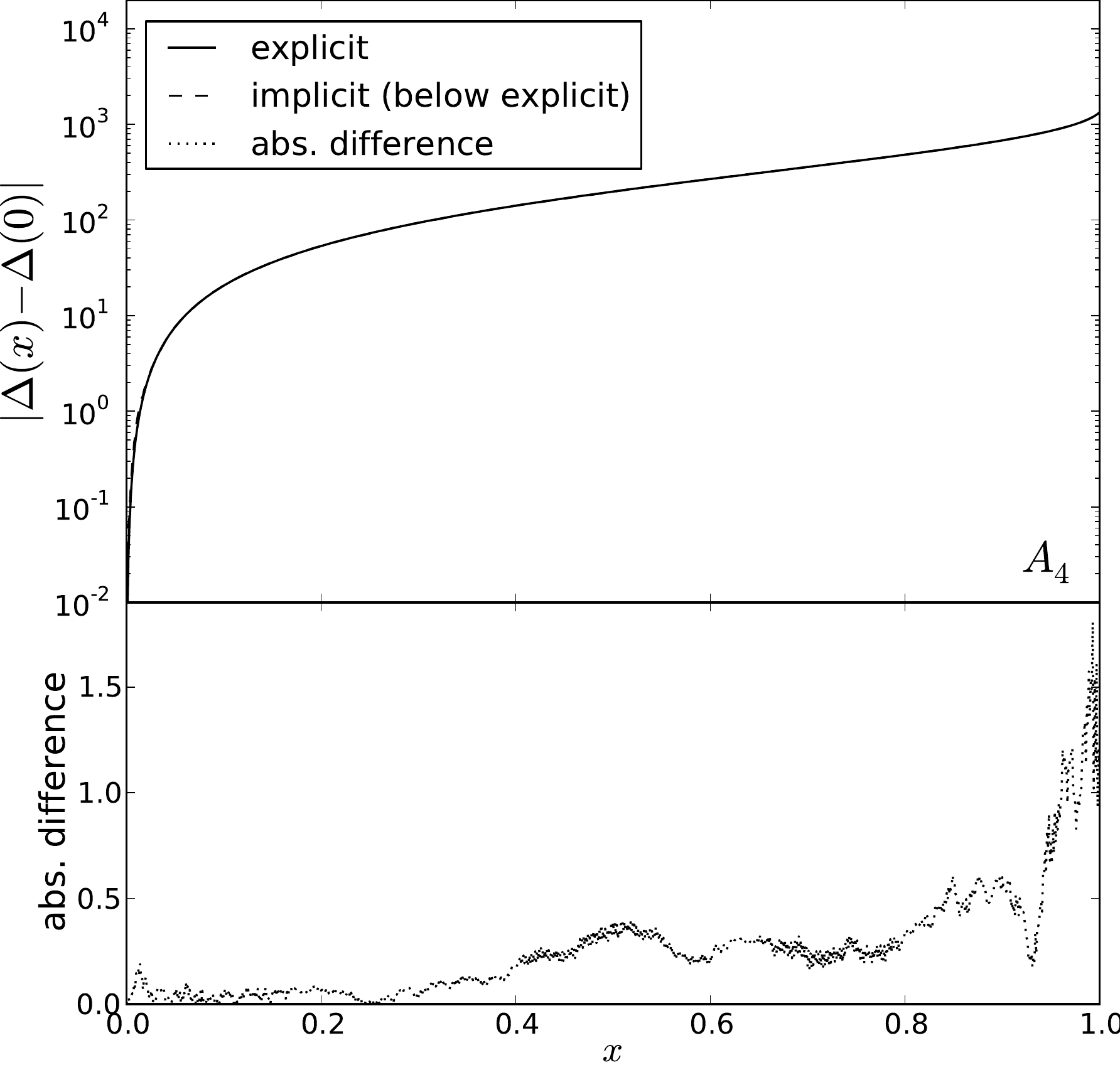}
\includegraphics[width=\columnwidth]{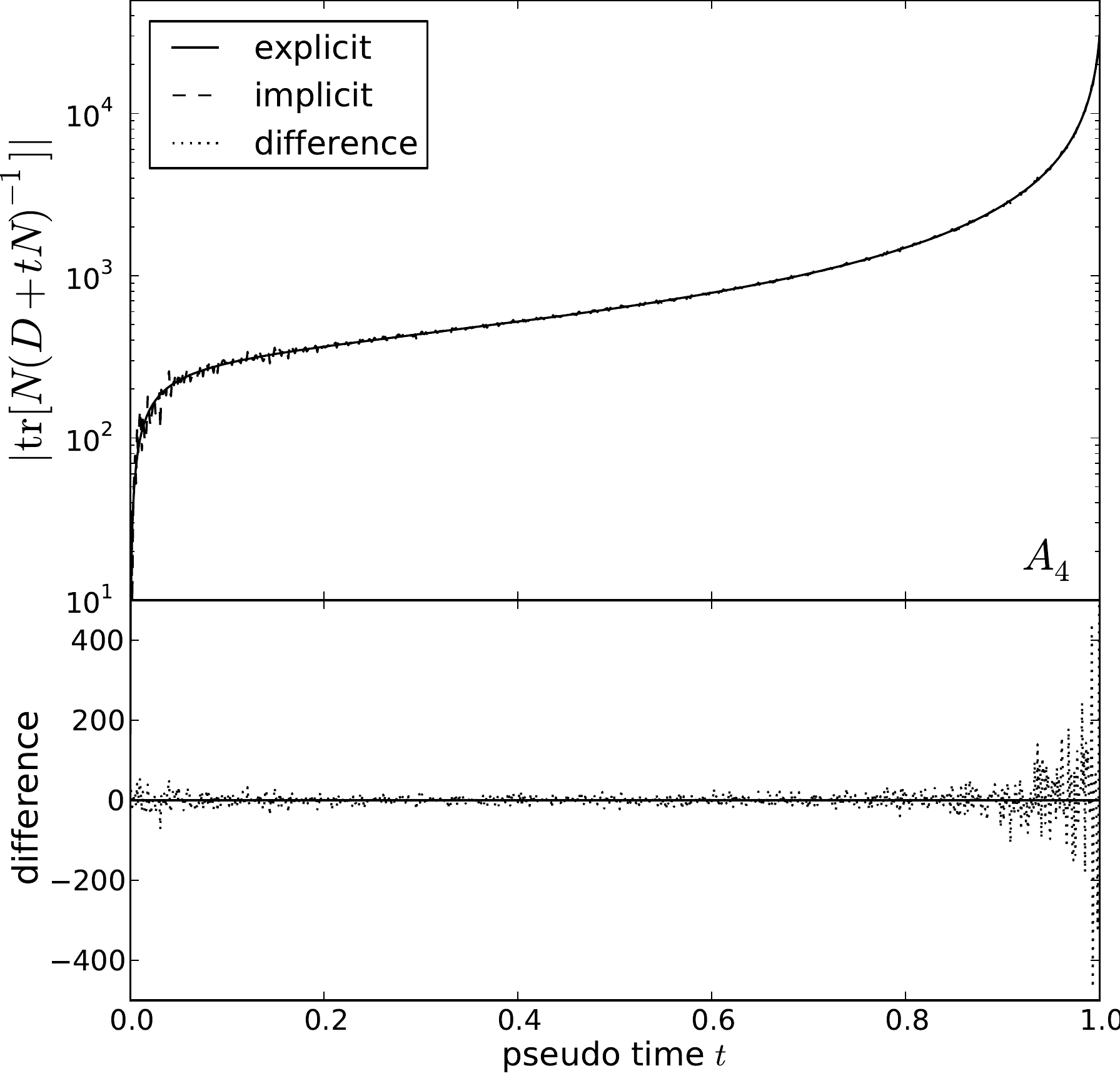}
\caption{The integrand of Eq.~\ref{eq:main_res} (lower panel) and $\Delta(x)$ (upper panel) for explicit and implicit representations of $A_4$.}%
\label{fig:matricies4}
\end{figure}

\begin{figure}[ht]
\includegraphics[width=\columnwidth]{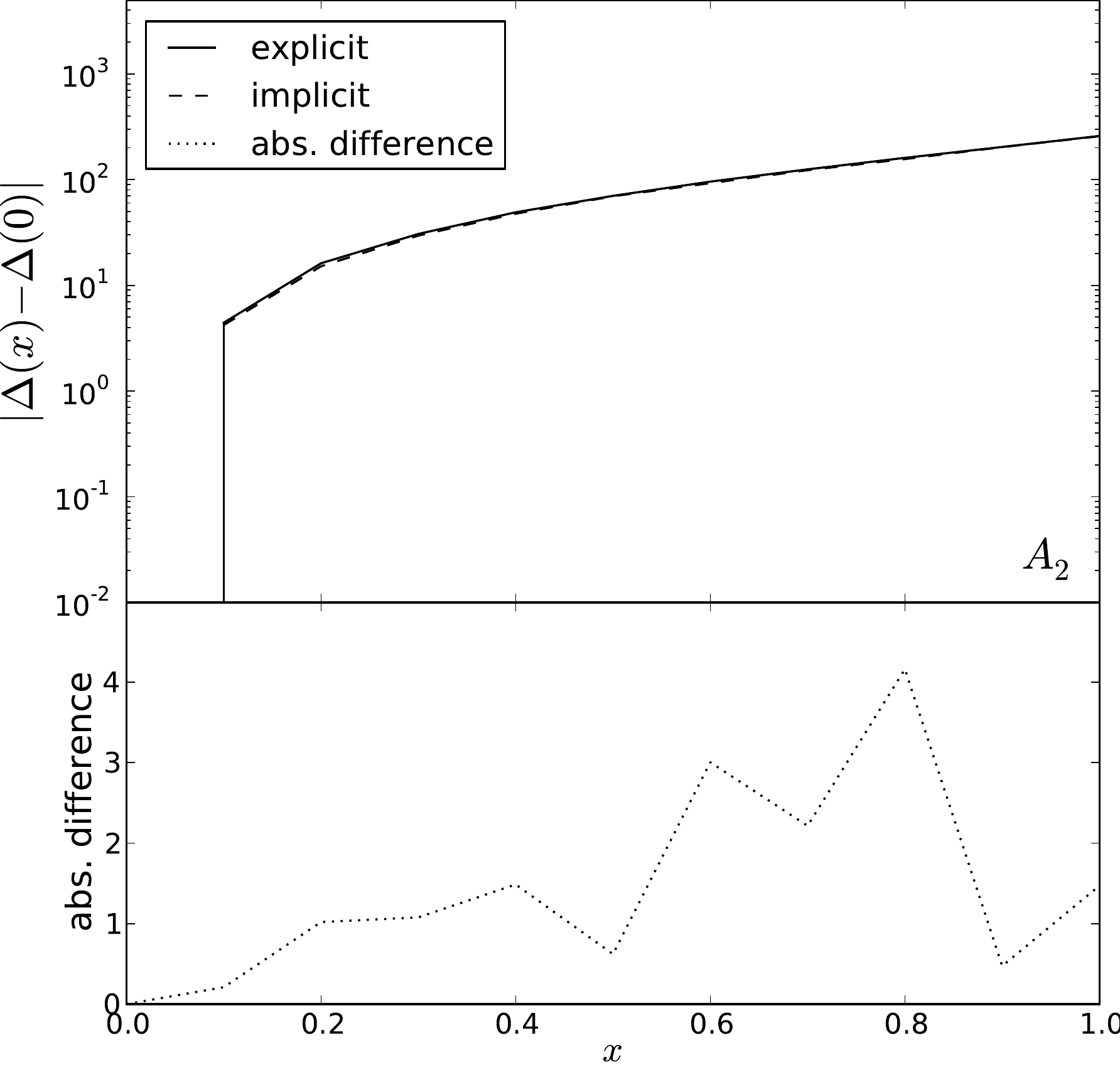}
\includegraphics[width=\columnwidth]{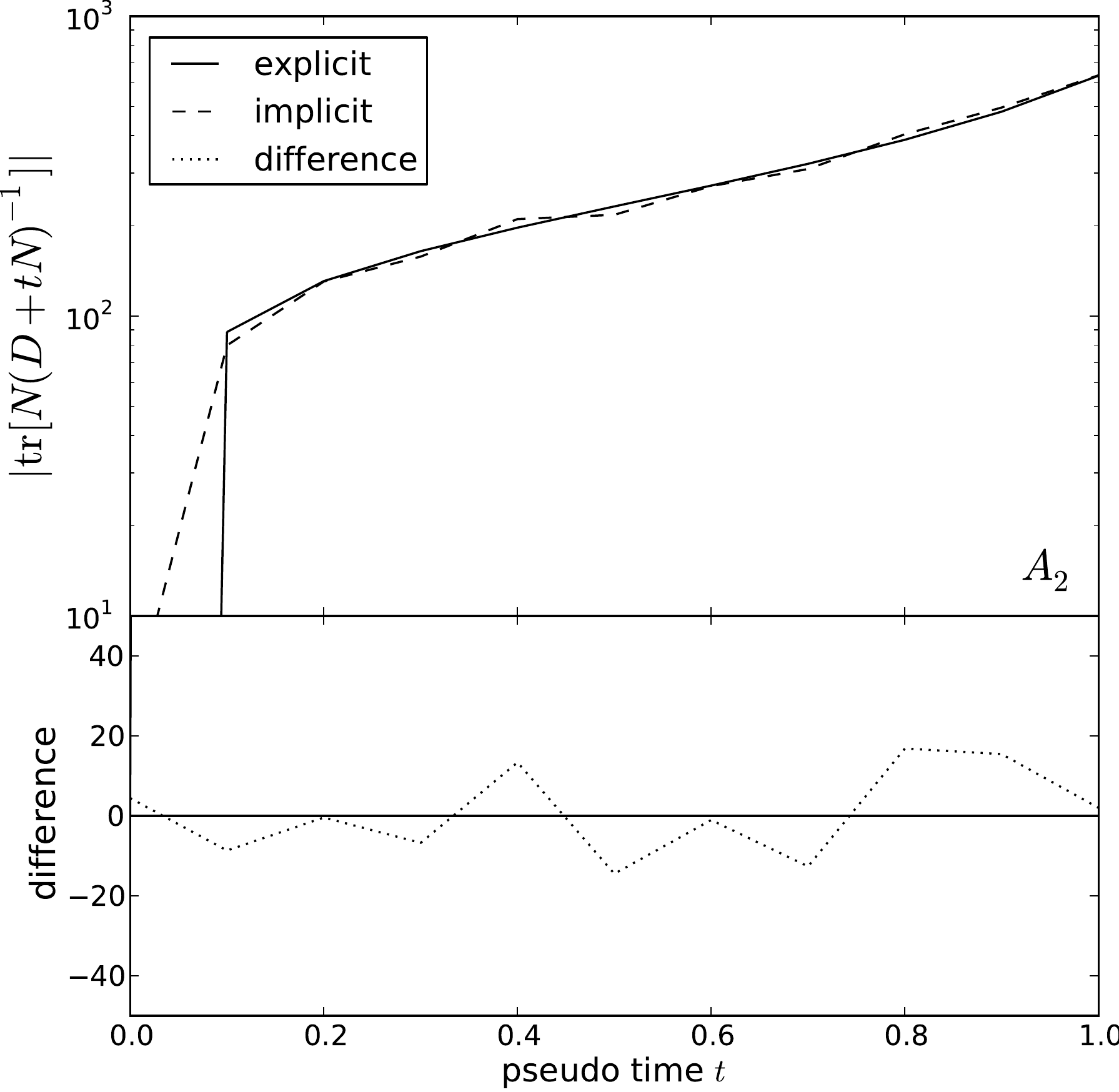}
\caption{The integrand of Eq.~\ref{eq:main_res} (lower panel) and $\Delta(x)$ (upper panel) for explicit and implicit representations of $A_2$ 
with only $m=10$ steps in pseudotime.}%
\label{fig:matricies2}
\end{figure}

\begin{figure}[ht]
\includegraphics[width=\columnwidth]{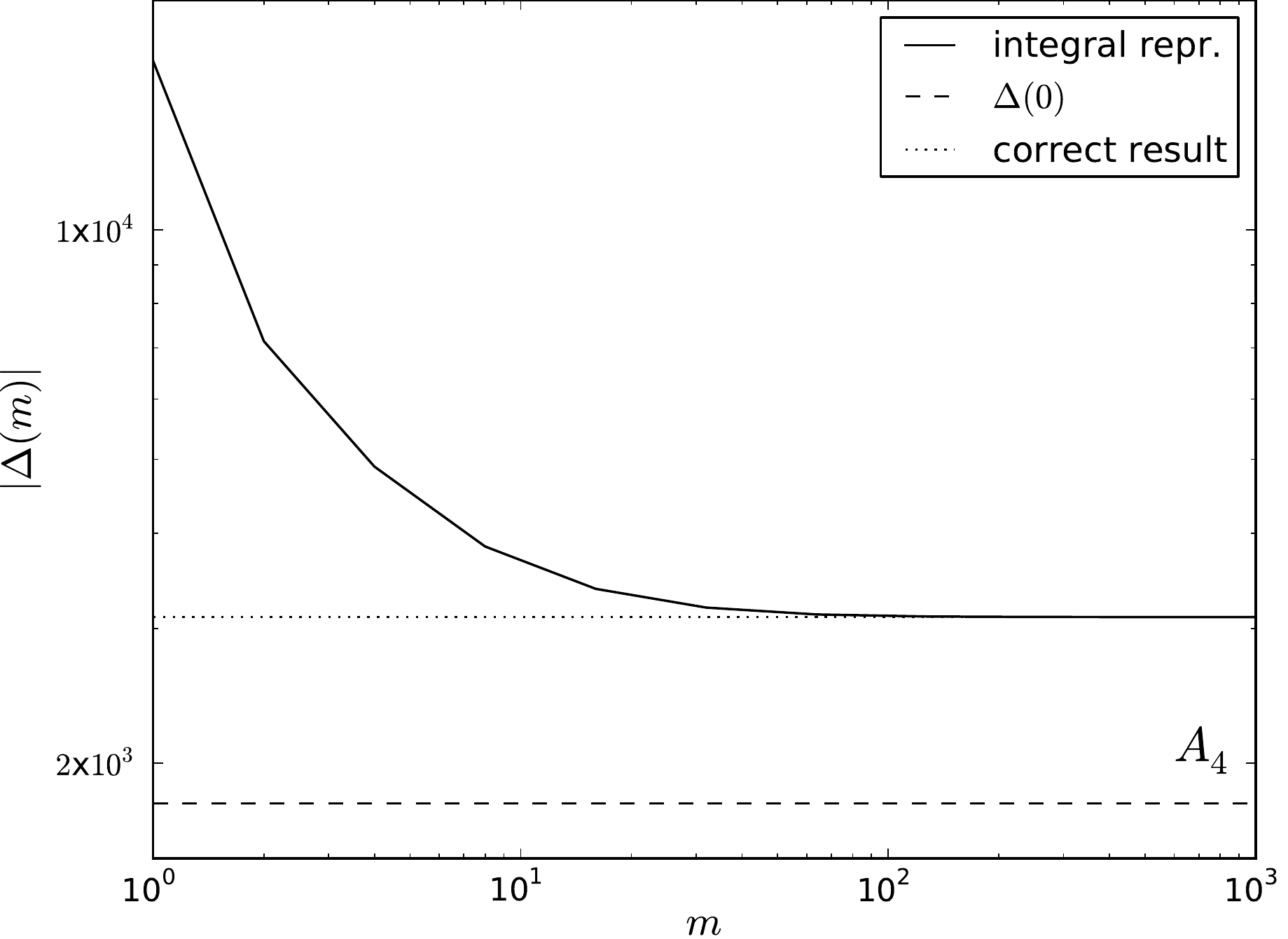}
\caption{Dependency of the determinant's result on the discretization of the integration interval into $m$ parts, using $A_4$. }%
\label{fig:m}
\end{figure}

After the separation of $A_{2}$ and $A_{4}$ into diagonal and off-diagonal parts by applying diagonal probing we calculate the integrands
of Eq.~(\ref{eq:main_res}) for the $m$-part-discretized interval of $t\in[0,1]$ by using the \texttt{conjugate gradient} method as well as
trace probing and perform the numerical integration afterwards by using \texttt{Simpson's} rule. The operator probing as well as the \texttt{conjugate gradient}
method have also been realized using \textsc{NIFTy}. Furthermore we introduce the quantities
\begin{equation}
\Delta(x) \equiv \int_0^x dt~ \mathrm{tr}\left[ N\left( D + tN \right)^{-1}\right] + \Delta(0),~~~x\in[0,1]
\end{equation}
to study the convergence to the final value and $\Delta(m)$ to investigate the dependency on the discretization of the integration 
interval, see Figs.~\ref{fig:matricies4}, \ref{fig:matricies2}, and \ref{fig:m}. 

We used a rather low sample size of $M=8$ for trace and diagonal probing [see Eqs.~(\ref{eq:trace}) and (\ref{eq:diag})] to demonstrate the
applicability of the method to large data sets.
The discretization of the pseudotime interval into $m$ parts was chosen to be $m=10^3$ for $A_4$ and only $m=10$ for $A_2$, 
see in particular Fig.~\ref{fig:m}, which illustrates the dependence of the probing result on $m$. 

\begin{table}[t]
\caption{\label{table1}
Results of the numerical determinant calculations with and without probing. The absolute errors of the probing method are defined 
by $\epsilon_1 = |\Delta_\mathrm{explicit}(1)-\Delta_\mathrm{implicit}(1)|$ and $\epsilon_2 = |\Delta_\mathrm{correct}-\Delta_\mathrm{implicit}(1)|$.
Differences between $\epsilon_1$ and $\epsilon_2$ arise from the discretized, numerical integration.}
\begin{ruledtabular}
\begin{tabular}{lrr}
 &  $A_2$ & $A_4$ \\
\colrule
$\Delta(0)$ & -1308.05  & -1771.57 \\
$\Delta_\mathrm{correct}$ & -1566.99  & -3107.28  \\
$\Delta_\mathrm{explicit}(1)$ & -1566.81  & -3107.29  \\
$\Delta_\mathrm{implicit}(1)$  &   -1565.33&  -3108.41\\
$m$  &   10&  1000\\
$M$  &   8&  8\\
$\epsilon_1$  & 1.48  &  1.12 \\
$\epsilon_2$  & 1.66  &  1.13 \\
\end{tabular}
\end{ruledtabular}
\end{table}

\subsection{Discussion}
The exact numerical values of the determinant calculation using explicit and implicit representations of $A_4$ and $A_2$ can be found in Table~\ref{table1}. The results
of the probing method (implicit) compared to the correct and the explicit method, where Eq.~(\ref{eq:main_res}) can be evaluated without using a \texttt{conjugate gradient}
or probing techniques, are accurate for both matrices. It is remarkable that despite using a relatively small sample size of $M=8$ the absolute errors 
remain relatively small. The reason for this is that the pseudotime integration over all probed integrands averages the probing error.
This is of particular importance when applying the log-determinant probing to large data sets, where a large sampling size should be avoided to save
computational time. 
These errors can be decreased further, of course, by an increase of the sampling size and a refinement of the numerical integration. 

The results of the trace (integrand) probing and the determinant's convergence behavior as well as their respective errors with respect to the explicit representation 
can be found in Figs.~\ref{fig:matricies4} and \ref{fig:matricies2}. Note that the scaling of the ordinate is logarithmic. For both matrices, but especially for $A_4$,
the largest contribution to the integral of Eq.~(\ref{eq:main_res}) comes from late $t$-values. Therefore, if dealing with big data sets,
one could divide the integration interval not into $m$ equal parts but by starting with a rather coarse discretization for small
$t$-values and subsequently refining it for larger values, e.g.,~by substituting $dt$ by $d\ln(t')$ and thereby saving computational costs.
This, however, might depend on the particular shape of the matrix and has to be studied case by case. 

The dependency of the numerical value of the determinant of $A_4$ on the discretization (in $m$ equal parts) of the integration interval
can be found in Fig.~\ref{fig:m} and shows that even a small value of $m$ adds significant corrections to the result. The result for $m\propto \mathcal{O}(10)$ is, for
instance, better than just using the determinant of the diagonal, $\Delta(0)$. This might be used in practice to investigate cheaply whether the 
nondiagonal structure of a matrix influences the determinant significantly.  

A huge advantage of the probing method discussed here is the possibility to parallelize the numerical calculation almost completely. 
To be precise, the diagonal probing beforehand, the pseudotime integral, as well as every single trace probing can be parallelized fully. 
The only operation that cannot be parallelized is the \texttt{conjugate gradient} method as it is a potential minimizer, 
using at least the previous step to calculate the next one.

The determination of a suitable choice of the involved parameters $m$ and $M$ as well as the precision parameters for the used \texttt{conjugate gradient} approach
and numerical integration method depend highly on the matrix to be studied. The computational costs and precision of the introduced determinant calculation
thus depend on the combination of the chosen methods for diagonal and trace probing, numerical
integration, the method to numerically invert the matrix $D + tN$, and the matrix $A$ itself.
Since it is therefore not possible to make general statements we consciously avoid here such a 
discussion of computational costs and precision with respect to $m$ and $M$. A more pragmatic way to estimate these parameters would be to 
downscale the problem of interest until the matrix of interest fits into the memory of the computer and to subsequently perform mock tests to obtain 
a suitable choice for the parameters discussed above. Afterwards these values can be extrapolated to the size of the real problem. 

\section{Applications in science}\label{sec:applications}
Within this section we present a selection of possible applications in science. Although there are a vast number of research fields and topics which might
benefit from the stochastic estimation of a log-determinant we focus henceforth on a selection of usages in Bayesian signal inference, in particular in
physics and only present simple examples. Exact, more complicated examples can be found in the cited works within this section.  

\subsection{Evidence calculations \& model selection}\label{sec:evidence}
The Bayesian evidence $\mathcal{P}(d)$ is a measure for the quality of the model and hence for all assumed model parameters for the data $d$ \cite{jaynes2003probability}.
To keep it short and simple we assume
a model that describes a linear measurement of a Gaussian signal $s$ with additive, signal-independent, Gaussian noise $n$, i.e.,
\begin{equation}
d = Rs +n,
\end{equation}
where $R$ represents a linear operator. A Gaussian distribution of a variable $x$ is defined by
\begin{equation}
\mathcal{P}(x)=\mathcal{G}(x,X)\equiv \frac{1}{\sqrt{|2\pi X|}}\exp\left\{ -\frac{1}{2}x^\dag X^{-1} x \right\}
\end{equation}
with related covariance matrix $X$ and mean 
\begin{equation}
 \left\langle x \right\rangle_{\mathcal{P}(x)} \equiv \int \mathcal{D}x~x~\mathcal{P}(x).
\end{equation}
$\int \mathcal{D}[\cdot]$ denotes a phase space integral and $|\cdot|$ the determinant. Under these circumstances the evidence can be calculated as
\begin{equation}\label{eq:evidence}
\begin{split}
\mathcal{P}(d) =& ~ \int \mathcal{D}s \int \mathcal{D}n ~\mathcal{P}(d,s,n)\\
		 =& ~ \int \mathcal{D}s \int \mathcal{D}n ~\delta(d -Rs - n)\mathcal{P}(n|s)\mathcal{P}(s)\\
		 =& ~ \sqrt{\frac{|2\pi C_{s|d}|}{|2\pi C_s|| 2\pi C_n |}}\exp\left\{-\frac{1}{2}\left(d^\dag C_n^{-1}d - j^\dag C_{s|d} j \right) \right\},
\end{split}
\end{equation}  
with 
\begin{equation}
 \begin{split}
 j=&~R^\dag C_n^{-1}d,\\ C^{-1}_{s|d}=&~R^\dag C_n^{-1}R +C_s^{-1},
 \end{split}
\end{equation}
and the signal and noise covariances $C_s$ and $C_n$, respectively. 
Therefore, to calculate the Bayesian model evidence, one often\footnote{By the word ``often'' we refer to cases, in which
at least one marginalization [see Eq.~(\ref{eq:marg})] can be performed analytically (approximated with high precision) to obtain a model-dependent determinant. }
has to calculate determinants of covariance matrices. 
This might be done by probing [Eq.~(\ref{eq:main_res})] if dealing with 
implicit matrices [last line of Eq.~(\ref{eq:evidence})] instead of performing the multidimensional integral 
[second last line in Eq.~(\ref{eq:evidence})] numerically. The latter has been done, for instance, in the
field of inflationary cosmology~\cite{2013arXiv1303.5082P,2014JCAP...03..039M} by the method of nested 
sampling~\cite{:/content/aip/proceeding/aipcp/10.1063/1.1835238,2009MNRAS.398.1601F}.   

This is especially of importance in the field of model selection or comparison \cite{jaynes2003probability}, 
where from an observation -- the data -- one wants to infer which theory
reproduces the observation best. Switching from one model to another means, for 
instance\footnote{We focus here on $R$ for simplicity only. One could also, additionally, exchange the 
prior covariances $C_n$ and $C_{s}$, the assumed prior statistics, the parametrization of the data, and so on. }, to 
exchange $R$ in Eq.~(\ref{eq:evidence}), which directly affects the determinant
containing $C_{s|d}$. Thus, the calculation of the determinant is mandatory here.  

\subsection{Posterior distribution including marginalizations}\label{sec:post}
In the field of signal inference one is typically interested in reconstructing a set of $i$ parameters $p^i$ with uncertainty from some observation, the data $d$.
This information is delivered by the posterior, given by\footnote{Note that in this case the evidence is just a scalar which normalizes the posterior, therefore
we merely state proportionalities.} \cite{Bayes01011763} $\mathcal{P}(p^i|d) \propto \mathcal{P}(d|p^i)\mathcal{P}(p^i)$. Often, however, this inference problem 
is degenerate, caused by a so-called nuisance 
parameter. For example, consider the 
calibration of an instrument is of interest and not the signal. In this case the signal $s$ represents
the nuisance parameter. The common procedure to circumvent this problem is to marginalize over these parameters,
\begin{equation}\label{eq:marg}
 \mathcal{P}(p^i|d) \propto \int \mathcal{D}s \int \mathcal{D}n ~ \mathcal{P}(d, s, n|p^i)\mathcal{P}(p^i).
\end{equation}
To continue with the simple example of Sec.~\ref{sec:evidence} we assume again Gaussian distributions for $s$ and $n$
and a linear measurement but with explicit dependency on $p^i$, i.e., $d = (Rs)[p^i] + n$. If we further follow the example of calibration,
the parameter $p^i$ might be a calibration coefficient, thus affecting only $R$. This yields $(Rs)[p^i]=R[p^i]s$ and therefore
\begin{equation}
 \mathcal{P}(p^i|d) \propto \left\{ \int \mathcal{D}s ~ \mathcal{G}\left(d - R[p^i]s ,C_n\right)\mathcal{G}(s,C_s)\right\}\mathcal{P}(p^i).
\end{equation}
This integration can be performed analytically,
~producing an in general non-Gaussian probability distribution with $p^i$-dependent normalization (and exponent) similar to Eq.~(\ref{eq:evidence}),
\begin{equation}\label{eq:marg_post}
\begin{split}
\mathcal{P}(p^i|d) \propto &~ \sqrt{\left|2\pi C_{s|d}[p^i]\right|}~\mathcal{P}(p^i) \\
		 &~\times \exp\left\{\frac{1}{2} j^\dag[p^i] C_{s|d}[p^i] j [p^i] \right\}
\end{split}
\end{equation} 
with $C_{s|d}[p^i]$ and $j[p^i]$ now containing $R[p^i]$ instead of $R$.
In case the covariance matrices or $R[p^i]$ are only given by a computer routine (implicit representation of a matrix) one could use Eq.~(\ref{eq:main_res})
to probe the determinant.

A variety of scientific fields are affected by this problem. For example, the extraction of the level of non-Gaussianity of the cosmic
microwave background \cite{paper1,2014JCAP...06..048D} in cosmology, the problem of self-calibration \cite{2002MNRAS3351193B,2014PhRvE..90d3301E,2015PhRvE..91a3311D} 
in general, or lensing in astronomy \cite{2003PhRvD..67d3001H}.

\subsection{Realistic astronomical example}
In order to study a more realistic example we consider a measurement device with spatially constant but unknown calibration amplitude, parametrized by 
$1+\gamma \in \mathds{R}$, scanning a specific patch of the sky. The measured and assumed to be Gaussian sky signal $s$ is affected by the
instrument via a convolution $\mathcal{C}$ with a Gaussian kernel of standard deviation $\sigma = 0.05$. 
Additionally, the observation might be disturbed by fore- and backgrounds. For this reason we include an observational mask $M_o$, which cuts out $20\%$ of the 
sky. The noise $n$ is still assumed to be Gaussian and uncorrelated with the signal. Hence, the measurement equation is given by
\begin{equation}
d=R\left[\gamma\right] s +n = (1+\gamma)  M_o \mathcal{C} s +n. 
\end{equation}
To calibrate the measurement device the calibration posterior $\mathcal{P}(\gamma|d)$ has to be determined. The resulting calibration mean 
$\left\langle \gamma \right\rangle_{\mathcal{P}(\gamma | d)}$ can be regarded as an external calibration if the \textit{a priori} knowledge on the signal is 
sufficiently strong. Otherwise one could infer the signal and calibration amplitude $\gamma$ simultaneously from data using iterative 
approaches~\cite{2014PhRvE..90d3301E}. Using Eq.~(\ref{eq:marg_post}) as well as a flat prior on $\gamma$ we obtain 
\begin{equation}\label{eq:astro}
\begin{split}
\ln \mathcal{P}(\gamma|d) =&~ -\frac{1}{2}\ln\left| C^{-1}_{s|d}\left[\gamma\right]\right| + \frac{1}{2} j^\dag\left[\gamma\right] C_{s|d}\left[\gamma\right] j\left[\gamma\right]\\
			    &~+ \mathrm{const}.,
\end{split}
\end{equation} 
which exhibits in particular the $\gamma$-dependent determinant
\begin{equation}\label{eq:astro_det}
\left|C^{-1}_{s|d}\left[\gamma\right]\right|= \left|(1+\gamma)~ \mathcal{C}^\dag M_o^\dag C_n^{-1} M_o \mathcal{C} (1+\gamma) +C_s^{-1}\right|. 
\end{equation}

\begin{figure}[ht]
\includegraphics[width=\columnwidth]{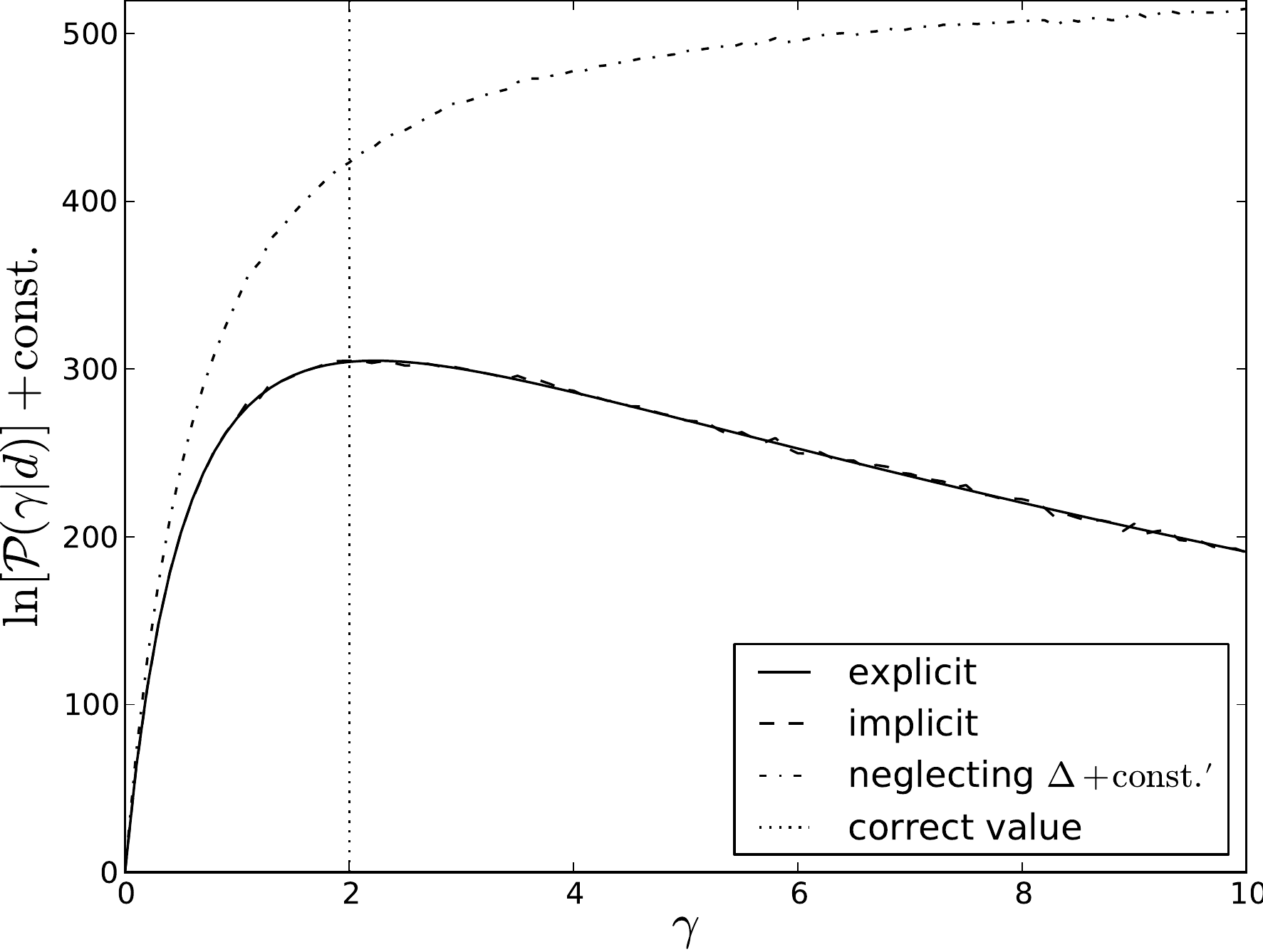}
\caption{Logarithmic posterior of the calibration amplitude parameter $\gamma$ using implicit and explicit representations of the involved 
operators, see Eq.~(\ref{eq:astro}) and Eq.~(\ref{eq:astro_det}) for details. The abbreviation $\Delta$ denotes the logarithm of the term given by Eq.~(\ref{eq:astro_det}).}%
\label{fig:expl}
\end{figure}

For the numerical evaluation of Eq.~(\ref{eq:astro}) we use the settings of Sec.~\ref{sec:num}
with $C_s(k,k') = \left(1 + k\right)^{-3}\delta_{kk'}$, a calibration amplitude parameter of $\gamma=2$, and a noise covariance of
$\left(C_n\right)_{x,x'} = 10^{-1}\delta_{xx'}$  to generate a data realization. 
The pseudotime interval has been discretized into $10^2$ parts. The numerically determined calibration posterior for a given data realization can be found in 
Fig.~\ref{fig:expl}, which demonstrates again the efficiency of the stochastic method using only eight probes for a single trace probing operation. The figure 
also illustrates the impact of the determinant on the log-posterior, which would not peak in the shown interval without it. 
\section{Summary}\label{sec:summary}
Motivated by the problem of finding a way to efficiently determine the determinant of an implicitly defined matrix or operator, we derived a formula,
Eq.~(\ref{eq:main_res}), representing a stochastic estimate of its log-determinant. This has been achieved by reformulating the log-determinant as an 
integral representation and transforming the involved terms into stochastic expressions, which includes a numerical integration and a trace probing.
Numerical examples have shown that the discretization of the integration interval may be very coarse in case the probed operator is sufficiently
diagonal. In case it exhibits a significant nondiagonal structure one has to fine-grain the discretization of this interval. The number of probes necessary for the trace probing,
however, remains very low in the studied examples. These facts combined with the almost complete parallelizability of this approach might keep 
the computational costs within reasonable limits in many situations.

This method clearly has more general applications but might in particular be useful for Bayesian signal inference and model comparison when dealing with 
large data sets as often given, for instance, in astronomy and cosmology. To be precise, it might be beneficial in all fields where the numerical calculation of a 
determinant of an operator is mandatory.

\begin{acknowledgments}
We gratefully acknowledge Maksim Greiner for discussions and David Butler for useful comments on the manuscript. 
All calculations were realized using \textsc{NIFTy} \cite{2013AA...554A..26S} 
to be found at \url{http://www.mpa-garching.mpg.de/ift/nifty}.
\end{acknowledgments}

\medskip

\appendix
\section{Integral representation of the log-determinant of a matrix}\label{app:integral}
Here Eq.~(\ref{eq:main_res}) is derived. Following Sec.~\ref{sec:main} the log-determinant $\Delta$ of an operator $A$ can be parametrized by
$\Delta = \ln[\mathrm{det}(D+N)]$ with $D$ being the diagonal and $N$ the off-diagonal part of $A$. Since $\Delta$ can be Taylor-expanded for 
small $N$ (spectrally compared to $D$) only, we employ a method from the field of renormalization theory \cite{2011PhRvD..83j5014E,2015PhRvE..91a3311D}.     
Accordingly, we introduce an expansion parameter $\delta t\ll1$ to suppress the influence of $N$. In particular, we replace $\Delta$ by 
$\ln[\mathrm{det}(D+\delta t N)]$ for a moment. For sufficiently small values of $\delta t$, in the following interpreted as tiny pseudotime steps, we can approximate $\Delta$ by Eq.~(\ref{eq:approx}). 
Theoretically, a single pseudotime step could be infinitesimal small, enabling the formal definition of the 
derivative
\begin{equation}
\begin{split}
\frac{d\Delta(t)}{dt} \equiv& \lim_{\delta t \rightarrow 0} \frac{\ln[\mathrm{det}(D+(t+\delta t) N)] - \ln[\mathrm{det}(D+ t N)]}{\delta t}\\
													 =& \lim_{\delta t \rightarrow 0}~ \frac{1}{\delta t} \ln\left[\mathrm{det}\left(\mathds{1} + \delta t N [D+t N]^{-1} \right)\right]\\
													 =& \lim_{\delta t \rightarrow 0}~ \frac{1}{\delta t} \mathrm{tr}\left[\ln\left(\mathds{1} + \delta t N [D+t N]^{-1} \right)\right]\\
													 =&~ \mathrm{tr}\left[ N [D+t N]^{-1} \right],
\end{split}
\end{equation}
with the definition
\begin{equation}
 \Delta(t) \equiv \ln[\mathrm{det}(D+t N)]. 
\end{equation}
Integrating the pseudotime derivative of $\Delta(t)$ yields the integral representation of the log-determinant,
\begin{equation}\label{app:main}
\Delta = \int_0^1 dt~ \mathrm{tr}\left[ N\left( D + tN \right)^{-1}\right] + \Delta(0).
\end{equation}
This integral representation has also been found by Ref.~\cite{du2005integral}, where its validity has been proven for weak diagonal dominant and Hermitian
positive definite matrices. In particular one has to ensure the existence of the inverse matrix of the integrand of Eq.~(\ref{app:main}).

Finally, we replace the trace by stochastic trace probing and perform the pseudotime integral by an numeric integration method.
This yields
\begin{equation}
\Delta 	= \int_0^1 dt~\left\langle \xi^\dag  N\left( D + tN \right)^{-1} \xi \right\rangle_{\{\xi\}} + \Delta(0).
\end{equation}

\bibliography{bibliography}

\end{document}